\documentclass[]{spie}  %>>> use for US letter paper
\usepackage{amsmath,amsfonts,amssymb,aas_macros}
\usepackage{graphicx}
\usepackage[colorlinks=true, allcolors=blue]{hyperref}

\newcommand{\nlhref}[1]{\href{#1}{\nolinkurl{#1}}} %automatically create exact url with href

\title{The Roman Coronagraph Community Participation Program: observation planning}

\author[a]{Schuyler G. Wolff}
\author[b]{Jason Wang}
\author[c]{Karl Stapelfeldt}
\author[d]{Vanessa P. Bailey}
\author[e]{Dmitry Savransky}
\author[a]{Justin Hom}
\author[f]{Beth Biller}
\author[g]{Wolfgang Brandner}
\author[a]{Ramya Anche}
\author[b]{Sarah Blunt}
\author[h]{Marah Brinjikji}
\author[i]{Julien H. Girard}
\author[j]{Oliver Krause}
\author[k]{Zhexing Li}
\author[l,m]{John Livingston}
\author[n]{Maxwell A. Millar-Blanchaer}
\author[b]{Malachi Noel}
\author[i]{Laurent Pueyo}
\author[o]{Robert J. De Rosa}
\author[j]{Matthias Samland}
\author[a]{Nicholas Schragal}
\affil[a]{Steward Observatory and the Department of Astronomy, The University of Arizona, 933 N Cherry Ave, Tucson, AZ, 85721, USA}
\affil[b]{Center for Interdisciplinary Exploration and Research in Astrophysics (CIERA) and Department of Physics and Astronomy, Northwestern University, Evanston, IL 60208, USA}
\affil[c]{Jet Propulsion Laboratory, California Institute of Technology, Mail Stop 321-100, 4800 Oak Grove Drive, Pasadena CA 91109 USA}
\affil[d]{Jet Propulsion Laboratory, California Institute of Technology, 4800 Oak Grove Drive, Pasadena CA 91109 USA}
\affil[e]{Sibley School of Mechanical and Aerospace Engineering, Cornell University, Ithaca, NY, 14853, USA}
\affil[f]{Institute for Astronomy, University of Edinburgh, EH9 3HJ, Edinburgh UK}
\affil[g]{Max Planck Institute for Astronomy, 69117 Heidelberg, Germany}
\affil[h]{School of Earth and Space Exploration, Arizona State University, 781 E Terrace Mall, Tempe, AZ 85287, USA}
\affil[i]{Space Telescope Science Institute, 3700 San Martin Drive, Baltimore, MD 21218, USA}
\affil[j]{Max Planck Institute for Astronomy, Koenigstuhl 17, 69117 Heidelberg, Germany}
\affil[k]{Department of Earth and Planetary Sciences, University of California, Riverside, CA 92521, USA}
\affil[l]{Astrobiology Center, NINS, 2-21-1 Osawa, Mitaka, Tokyo 181-8588, Japan}
\affil[m]{National Astronomical Observatory of Japan, NINS, 2-21-1 Osawa, Mitaka, Tokyo 181-8588, Japan}
\affil[n]{University of California, Santa Barbara, California, 93106, USA}
\affil[o]{European Southern Observatory, Alonso de C\'{o}rdova 3107, Vitacura, Santiago, Chile}

\begin{document} 
\maketitle

\begin{abstract}
The Coronagraph Instrument onboard the Nancy Grace Roman Space Telescope is an important stepping stone towards the characterization of habitable, rocky exoplanets. In a Observation Phase conducted during the first 18 months of the mission (expected to launch in late 2026), novel starlight suppression technology may enable direct imaging of a Jupiter analog in reflected light. Here we summarize the current activities of the Observation Planning working group formed as part of the Community Participation Program. This working group is responsible for target selection and observation planning of both science and calibration targets in the technology demonstration phase of the Roman Coronagraph. We will discuss the ongoing efforts to expand target and reference catalogs, and to model astrophysical targets (exoplanets and circumstellar disks) within the Coronagraph’s expected sensitivity. We will also present preparatory observations of high priority targets. 
\end{abstract}

% Include a list of keywords after the abstract 

\section{INTRODUCTION}
\label{sec:intro}  % \label{} allows reference to this section

The Coronagraph Instrument onboard the Nancy Grace Roman Space Telescope (scheduled to launch no later than May 2027) will test and advance key high contrast imaging technologies in space \cite{2020arXiv200805624M,2020SPIE11443E..1UK}.   
The Roman Coronagraph is allocated 90 days in the first 18 months of operations with a requirement to demonstrate a contrast ratio of $1.1 \times$ 10$^{-7}$ in the visible at 6 to 9 $\lambda$/D ($0.3^{\prime\prime}-0.45^{\prime\prime}$) around a bright star (TTR5)\cite{2023SPIE12680E..0TB}. 
The achieved performance is expected to be even better with potential contrasts of a few $\times$ 10$^{-9}$ and a field of view ranging from 3 to 20 $\lambda$/D using various imaging modes. 

This will enable optical photometry of known self-luminous planets, attempts at the first detections of known radial velocity planets in reflected light, opportunistic searches for exozodiacal light in nearby sun-like stars \cite{2022PASP..134b4402D}, and observations (including polarimetry) of known warm debris disk systems.

The unique classification of the Coronagraph Instrument as a technology demonstration rather than a science instrument has motivated the creation of a unique support team structure. A Community Participation Program (CPP) solicited principal investigators from the major US and international partners to aid the Coronagraph Project Team and the Science Support Center to maximize the long-term value of the technology demonstration activities and datasets to the wider community.
Companion SPIE proceedings introduce the structure and goals of the CPP and its working groups \cite{spie_cpp, spie_drp}. Here we summarize the activities of one of the four Working Groups within the CPP; the Observation Planning Working Group. 

The Observation Planning Working Group (OPWG) is responsible for target selection (for both science and calibration targets)\cite{2022SPIE12180E..1ZZ}, target database development and maintenance, precursor observations, reference star identification, development and maintenance of the exposure time calculator, astrophysical modeling, and observation planning for both the baseline performance demonstration and additional observing modes \cite{2021SPIE11823E..1YR}.
The team includes $\sim$ 50 members across all career stages and international participation from ESA, JAXA, CNES and the Max Planck Institute for Astronomy. The first meeting was held in November, 2023 with early efforts focused on target vetting observations. 

Precursor observations of the fields of interest for several high priority Roman Coronagraph science targets are presented in Section \ref{sec:precursor}. Section \ref{sec:referencestars} details the efforts to develop a catalog of wavefront reference stars used for coronagraphic dark hole maintenance. Section \ref{sec:tools} provides updates to software tools in aid of observation planning. Lastly, opportunities for involvement by interested community members are presented in Section \ref{sec:community}.

\section{Precursor Observations of Target Stars}
\label{sec:precursor}
During Roman Coronagraph's nominal 2200~hr of observing time, it will likely only have time for deep imaging and spectroscopy of 1-2 cold Jupiter analogs in reflected light. It is imperative that we choose targets wisely and vet them thoroughly. Although the planets themselves are all confirmed via radial velocity, these systems have never been imaged before; hence, faint background stars are still potential contaminants. Contaminants come in two flavors: ``interlopers'' falling in the field of view and nearby ``glint stars'' that scatter light into the field of view. We obtained Keck/NIRC2 precursor imaging of several systems in 2019-2021 to mitigate this risk.

Although Roman Coronagraph will have orders of magnitude better contrast performance, NIRC2 precursor imaging taken well ahead of time can still be useful. The reason is twofold: (1) dust extinction from background stars is substantially weaker at near infrared (NIR) than at visible and (2) many CGI target stars have large proper motions ($>300$ mas/yr). We are therefore able to achieve meaningful sensitivity to background contaminants at the locations where the stars will be in 2027-2028, when Roman will observe. 

We ran simulations of background star contamination rates, based on Roman Coronagraph detection limit predictions and on Besancon Galactic stellar population models\cite{2014A&A...564A.102C}. Prior work cross-checked Besancon predictions against Gaia detections in the vicinity of our targets of interest, showing agreement to better than 25\% for all targets except the one with Galactic Latitude of  only 1 degree, where background source predictions are most sensitive to dust extinction models\cite{2021arXiv211008097C}. 

We used published NIRC2 contrast curves\cite{2015ApJS..216....7B}, and extrapolated to a conservative background limit of H=23 to predict the fraction of background star contaminants that NIRC2 could detect. In 2019-2021, we observed 4 known RV planet systems where NIRC2 was predicted to have meaningful sensitivity to background contaminants:  HD~190360, HD~134987, HD~154345, and $\upsilon$~And. 

We have developed a data reduction pipeline to process the NIRC2 data that is available on Github \footnote{\nlhref{https://github.com/MalachiNoel3/kecknirc2pipeline}}. The pipeline performs basic processing steps such as bad pixel detection and masking, dark subtraction, flat field creation and correction, and distortion correction using the distortion solution from \citenum{Service2016}. It performs relative image alignment (to align the speckles) by cross correlating the speckle field of each exposure relative to a chosen reference image. The absolute star center is found using the radon-transform-based approach \cite{Wang2014,Pueyo2015} implemented in \texttt{pyKLIP} \cite{Wang2015}. This technique uses the radially elongated speckles to locate the star center. Particular care is needed to handle the fact the star is placed near the edge of the detector in order to look at the field where the star will be during the nominal Coronagraph Instrument observation phase. We pad all images with 550 pixels of NaNs on each side to expand the size of the total image. This way, when the image is rotated about the star with North pointing up at the end in \texttt{pyKLIP}, the rotated field of view is still in the image. 

After pre-processing, the images of each star is fed into the NIRC2 interface of \texttt{pyKLIP}. The Karhunen–Lo\`eve Image Projection (KLIP)  algorithm \cite{Soummer2012} is used in conjunction with angular differential imaging \cite{Marois2006}. The images are then rotated about the star so that North is pointed up and stacked together using median combination. Candidate background objects are then identified by eye. The next steps are to uniformly process all data taken so far and measure the properties of the sources in the field to determine possible background objects in the field of view of the Coronagraph Instrument. This work is ongoing.

\section{Towards A Reference Star Catalog}
\label{sec:referencestars}

Both Roman and the future Habitable Worlds Observatory will rely on a catalog of dedicated reference stars to ``dig a dark hole''; suppress stellar speckle noise and provide a dark, high contrast region close-in to the central star \cite{2022JATIS...8a9002P}. 
This requires active control of the incoming wavefront using on board deformable mirrors to achieve contrasts of $10^{-7}$ or better.  The dark hole `digging' is performed using a bright reference star, and these deformable mirror settings are applied to the target star. To deliver the expected contrast in a reasonable amount of time, reference stars must be bright (Vmag $< 3$ provides a dark hole digging time $< 17$ hrs) single star systems. Moreover, it has been shown in detailed modeling that telescope pointing jitter, or equivalently resolved structure of even 1-2 mas can impact the contrast achievable in the dark hole \cite{2023JATIS...9d5002K}.
Thus, these wavefront reference stars are required to have a diameter of $<$ 2 mas, which induces a factor of only 1.5 degradation in contrast, and must be single stars. 

\begin{figure}
    \includegraphics[trim={10cm 0 10cm 0},clip,width=\textwidth]{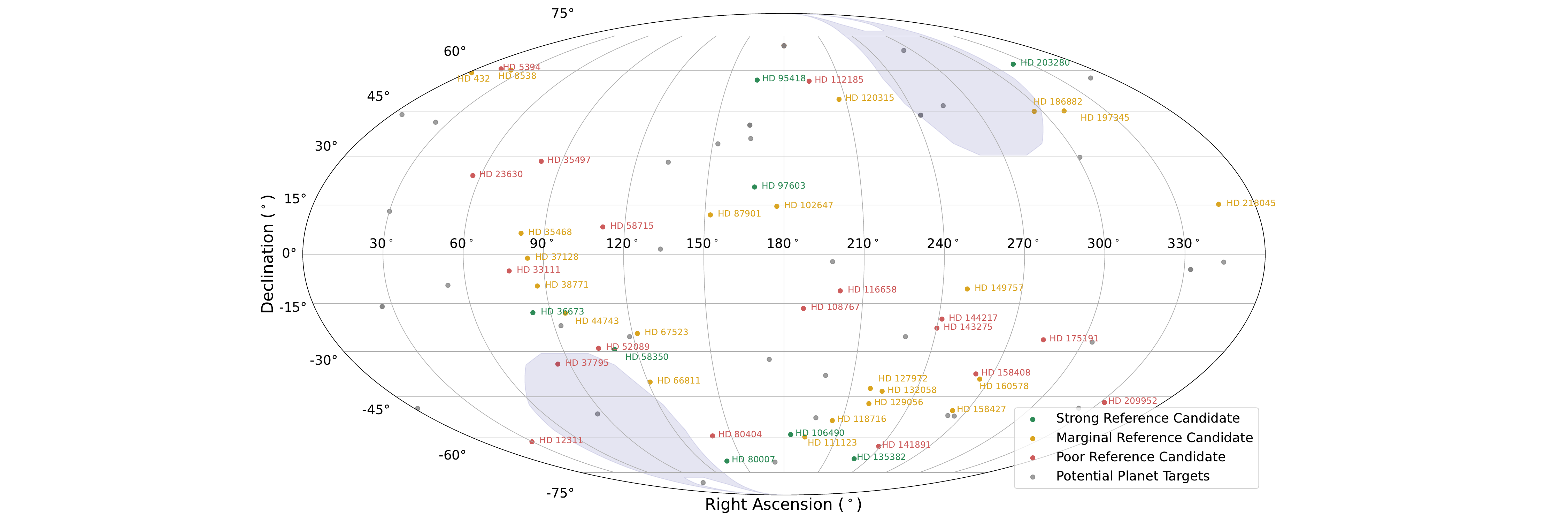}
    \caption{The on-sky distribution of candidate wavefront reference stars compared to the positions of potential planet targets (from the Imaging Mission Database, see Section \ref{sec:database}). Continuous Viewing Zones centered on the Ecliptic poles are displayed in blue. The wavefront reference sources are ranked according to their likelihood of hosting a companion that would degrade the contrast of the Roman Coronagraph. 
    }
    \label{fig:ref}
\end{figure}

Wavefront reference stars also need to be matched with Science targets in telescope sun pitch angle within a few degrees to maintain the thermal environment between the dark hole digging and target observations. The stellar diameter and brightness constraints significantly limit the number of sources in the sky that can be used as reference stars. 
To build a list of candidate reference stars, several stellar diameter catalogs were cross referenced \cite{2016yCat.2345....0D,2017yCat.2346....0B, 2017yCat..18330244S, 2017yCat..51530016S}, the resultant target list was vetted for known binaries using the WDS \cite{2001AJ....122.3466M} and SBC9 \cite{2004A&A...424..727P} catalogs, and cross-referenced for Gaia astrometric accelerations \cite{2019A&A...623A..72K,2022A&A...657A...7K}. This resulted in a total of $\sim$ 60 targets with a spatial distribution presented in Figure \ref{fig:ref}.

The stellar diameter and brightness constraint biases this sample towards distant, luminous stars. For a fixed apparent magnitude the stellar diameter $D \propto \sqrt{1/T}$ in the Rayleigh Jeans regime where the surface brightness scales with Temperature (T). Thus we require hot stars, and B-type are far more frequent in the solar neighborhood than O-type stars. The binary fraction for B-type stars is $>$ 90\%, and we anticipate companion detections around many of the candidate wavefront reference stars. 

Beginning in the 2024B semester, the Observation Planning Working Group has had several proposals accepted to began a companion vetting observing campaign. 
We combine interferometric measurements (capable of resolving close binaries within 100 mas) with speckle imaging (intermediate contrasts at 100 - 1200 mas) and ground-based high contrast imaging (outside of $\sim 20-40$ mas in J and H band, respectively with improved sensitivity to companions further out).
%We combine speckle imaging (probing companions at separations from 0.1 - 1.2$^{\prime\prime}$ ) with interferometric measurements (inside of 100 mas), and ground-based high contrast imaging observations (1 - 10$^{\prime\prime}$) for a complete companion survey. 
The interferometry observing campaign will also confirm/update existing stellar diameter measurements in a uniform manner. 
Unlike the target precursor observations described above, we are more concerned with bound companions for the wavefront reference stars. These sources also have low proper motion, and so cannot benefit from observations well in advance of launch to screen the locations where the stars \textit{will} be in 2027 for background objects.

Identifying bright targets with close binary companions may also present a technical demonstration opportunity. Characterizing on-sky performance of the Roman Coronagraph over a range of target properties, including stellar diameters and the presence, brightness, and separation of binaries, improves our collective understanding and predictive capabilities of coronagraph performance models.  In addition, a well-characterized catalog of wavefront reference stars both sets up the Roman Coronagraph for greatest success and serves as a starting point for the reference star catalog eventually needed for the Habitable Worlds Observatory.

The full catalog of wavefront reference candidate stars is publicly available\footnote{\nlhref{https://docs.google.com/spreadsheets/d/1p5r0VmjBCjXU25daJl5oJOPoPh1V79nuESbnwmca0s0/edit?usp=sharing}}. We encourage interested community members with access to relevant telescope facilities to become involved in the Reference target vetting process.

%A figure showing the sensitivity to companions enabled by our three main observing proposal types compared to CGI requirement? 

%A figure showing the spatial distribution of Reference stars against a background of science targets? Color code the reference stars based on how confident we are they are single. This may be more useful later on. 

\section{Updates to Existing Tools}
\label{sec:tools}

The CPP team leverages software toolkits previously developed by collaborators on the Roman Coronagraph Instrument teams and by the now disbanded Nancy Grace Roman Space Telescope Coronagraph Instrument Science Investigation Teams (SIT).\footnote{The full list of publications, data products, and software tools produced by the SIT can be accessed from \nlhref{https://romancgi.sioslab.com} and \nlhref{https://www.exoplanetdatachallenge.com/SIT}} For a review of previously implemented simulation and instrument performance tools see \cite{2020SPIE11443E..38D}. 
Updates to data simulation tools are being performed in tandem and with the support of the Observation Planning Working Group efforts (e.g. polarization of debris disks  \cite{2023PASP..135l5001A}). 
Here we present current and planned updates to existing software tools, with a focus on the observation planning for both the primary observing mode (the Hybrid Lyot Coronagraph centered at 575 nm enables imaging from 0.14$^{\prime\prime}$ – 0.44$^{\prime\prime}$, hereafter Band 1) and the additional wide field imaging, polarimetry and spectroscopy modes. 

\subsection{Imaging Mission Database}
\label{sec:database}

The Imaging Mission Database\footnote{\url{https://plandb.sioslab.com/}} collects information from the NASA Exoplanet Science Institute Exoplanet Archive (star, planet properties, orbital information). Following updates to the orbital fits, these data are combined with planet photometry\cite{batalha2018color} to predict the observability of known exoplanets with future high contrast imaging instruments. 
For a given planet, the $\Delta mag$ from the host star and projected separation is computed over the span of its orbit, over the range of unknown inclinations and values allowed for the other orbital elements by the RV data. 
Two-dimensional frequency maps of the joint distribution of planet projected separation and brightness are generated by sampling from the uncertainty distributions in the orbital parameters. 
For a full description see \url{https://plandb.sioslab.com/docs/html/index.html} and references therein. 
The Observation Planning Working Group has adopted the Imaging Mission Database as its primary target and reference database.

There are several planned upgrades for the Imaging Mission Database. 
The target database will be expanded to include non-planet hosting science targets, calibration sources, and wavefront reference stars. The inclusion of wavefront reference stars will enable quick and efficient selection of the best science/reference pairs to minimize slew times and thermal variations. 
Currently, the orbital information for each planet is taken from the default paremeter set solution available in the NASA Exoplanet Archive (hosted by IPAC). 
In cases where multiple radial velocity datasets exist, a comprehensive and robust orbital solution will be determined using the \texttt{Orbitize!} \cite{2020AJ....159...89B} and \texttt{RadVel} \cite{2018PASP..130d4504F} tools led by S. Blunt. 
Future Gaia DR4 exoplanet detections may also be viable Roman Coronagraph targets and will be integrated into the database. 
Finally, we plan to incorporate estimates of the Roman orbital ephemera to build an optimization scheme for science/wavefront reference pair selection.

\subsection{Exposure Time Calculator}

The Roman Coronagraph Exposure Time Calculator (ETC: \url{https://github.com/hsergi/Roman_Coronagraph_ETC}) is the publically accessible interface to the science camera of the Roman Coronagraph. It  presents two performance scenarios: an ``optimistic'' and a ``conservative'' case for the achieved contrast with instrumental and modeling parameters as described in \cite{10.1117/1.JATIS.6.2.027001, 10.1117/1.JATIS.2.1.011006, 10.1117/1.JATIS.6.3.039002, 2017ascl.soft06010S, 10.1117/12.2575983}.
Following the conclusion of the instrument-level functional, environmental, and performance tests \cite{spie_tvac} the performance scenarios are now being updated to represent the predicted on-sky performance. 

\section{Towards a Suite of Observing Programs}

The Roman Coronagraph will necessarily carry out a set of commissioning, calibration, and performance demonstration observations in support of the instrument's single TTR5 requirement of achieving 10$^{-7}$ contrast at a 6 $\lambda/D$ separation from a $V=5$ mag star.  Developing the set of observing programs needed to achieve this is a goal of the CPP OPWG over the next year.  But beyond this, the Roman Coronagraph has significant potential to act as a testbed for design issues being faced by the Habitable Worlds Observatory (HWO).

To capture a range of ideas for Roman Coronagraph Observations beyond TTR5, the OPWG plans to collect a set of observing program ideas that can be developed and then prioritized versus each other as Roman approaches launch. Each suggested program would be documented in a way that facilitates comparisons of cost in observing time and benefit to Roman Coronagraph objectives and HWO needs.  Observing program ideas that could be accomplished using similar datasets would be merged.  Observing programs on topics such as the stability of the instrument and telescope system, model validation, spectroscopy performance, polarimetry performance, low-order wavefront sensor performance, and extended source imaging - all of which go beyond Roman coronagraph's threshold requirement -  might be pursued.  Suggestions from the broader community will be solicited.  From the prioritized list of fully-defined observing programs going beyond TTR5, the intent would be to execute as many of them as possible during the instrument observation phase time allocation.

If the Roman Coronagraph is to be used to conduct additional observations beyond the initial 2200~hr observation phase - such as surveys of self-luminous exoplanets, radial velocity planets in reflected light, exozodiacal dust in nearby habitable zones, and/or warm debris disks, or other technology demonstrations or science observations  - then these programs need to be defined in terms of their cost in observing time, cost in terms of data analysis effort, and benefits to HWO architecture topics and the exoplanet and disk science communities.  Roughly 6-9 months after launch the Roman Project and NASA HQ plan to hold a review on whether to continue operations of the instrument beyond the first 18 months of the Roman mission.  Having fully-defined options for compelling science and technology programs - possibly multiple versions of each, to reflect the range of instrument performance that might be encountered on orbit - will facilitate this important decision on the future utilization of the instrument.  In the year before launch, working with the community, the CPP and OPWG expect to pursue development of these possible programs.

\section{Community Involvement}
\label{sec:community}

One of the primary mandates of the Roman Coronagraph Community Participation Program is to engage with the broader community on how
best to use the Coronagraph Instrument during its technology demonstration phase. 
To that end, the Obs. Planning WG is constructing a community interest survey to help prioritize the science and technology demonstration objectives that CPP will work to implement.  Participants will be asked to rank various science cases (e.g. imaging giant planets in reflected light, spectroscopy of a reflected light planet, polarization of exozodiacal light) and coronagraph performance trials (e.g. thermal stability, impact of stellar properties on achieved contrast, test wavefront control algorithms). Short answer questions will allow participants to recommend additional science and performance demonstration activities.   
In order to reach as diverse an audience as possible, the survey will be anonymous and will be accompanied by a Roman Coronagraph Pocket Guide to provide sufficient details on its capabilities to allow early career researchers to participate without an extensive literature search. 

\acknowledgments % equivalent to \section*{ACKNOWLEDGMENTS}       

This material is based upon work supported by NASA under award No. 80NSSC24K0217 and 80NSSC24K0087. 
This research has made use of the Imaging Mission Database, which is operated by the Space Imaging and Optical Systems Lab at Cornell University. The database includes content from the NASA Exoplanet Archive, which is operated by the California Institute of Technology, under contract with the National Aeronautics and Space Administration under the Exoplanet Exploration Program, and from the SIMBAD database, operated at CDS, Strasbourg, France. 
This research has made use of the NASA Exoplanet Archive, which is operated by the California Institute of Technology, under contract with the National Aeronautics and Space Administration under the Exoplanet Exploration Program.
We thank Tom Esposito for developing the NIRC2 interface for \texttt{pyKLIP} that was used in this work. 
 
%This unnumbered section is used to identify those who have aided the authors in understanding or accomplishing the work presented and to acknowledge sources of funding.  

% References
\bibliography{main} % bibliography data in report.bib
\bibliographystyle{spiebib} % makes bibtex use spiebib.bst

\end{document}